\documentclass[a4paper,11pt]{article}
\pdfoutput=1 

\usepackage{jcappub} 

\usepackage[T1]{fontenc} 
\usepackage{ifthen}
\usepackage{epsfig}
\usepackage{booktabs} 
\usepackage{natbib}
\usepackage{bbold}
\usepackage{diagbox}
\usepackage{gensymb}

\newcommand{\erf}{\mathop{\mathrm{erf}}}
\newcommand{\beqra}{\begin{flalign}}
\newcommand{\eeqra}{\end{flalign}}
\newcommand{\beq}{\begin{equation}}
\newcommand{\eeq}{\end{equation}}

\usepackage{ccicons} 
\usepackage{listings}
\usepackage{xcolor}   
\usepackage{braket}  
\usepackage{hyperref}
\usepackage{dsfont}  
\usepackage{leftidx} 
\RequirePackage{mathrsfs}
\RequirePackage{dsfont}
\usepackage{bm}
\usepackage{placeins}
\usepackage{float}
\allowdisplaybreaks

\title{\boldmath Projected sensitivity to sub-GeV dark matter of next-generation semiconductor detectors}
\author[a]{Erik Andersson}
\author[a]{Alex B\"okmark}
\author[a]{Riccardo Catena}
\author[a]{Timon Emken}
\author[a]{Henrik Klein Moberg}
\author[a]{and Emil \AA strand}
\affiliation[a]{Chalmers University of Technology, Department of Physics, SE-412 96 G\"oteborg, Sweden}
\emailAdd{erianb@student.chalmers.se}
\emailAdd{bokmark@student.chalmers.se}
\emailAdd{catena@chalmers.se}
\emailAdd{emken@chalmers.se}
\emailAdd{hmoberg@student.chalmers.se}
\emailAdd{asemil@student.chalmers.se}

\abstract{We compute the projected sensitivity to dark matter (DM) particles in the sub-GeV mass range of future direct detection experiments using germanium and silicon semiconductor targets.~We perform this calculation within the dark photon model for DM-electron interactions using the likelihood ratio as a test statistic, Monte Carlo simulations, and background models that we extract from recent experimental data.~We present our results in terms of DM-electron scattering cross section values required to reject the background only hypothesis in favour of the background plus DM signal hypothesis with a statistical significance, $\mathcal{Z}$, corresponding to 3 or 5 standard deviations.~We also test the stability of our conclusions under changes in the astrophysical parameters governing the local space and velocity distribution of DM in the Milky Way.~In the best-case scenario, when a high-voltage germanium detector with an exposure of $50$~kg-year and a CCD silicon detector with an exposure of $1$~kg-year and a dark current rate of $1\times10^{-7}$~counts/pixel/day have simultaneously reported a DM signal, we find that the smallest cross section value compatible with $\mathcal{Z}=3$ ($\mathcal{Z}=5$) is about $8\times10^{-42}$~cm$^2$ ($1\times10^{-41}$~cm$^2$) for contact interactions, and $4\times10^{-41}$~cm$^2$ ($7\times10^{-41}$~cm$^2$) for long-range interactions.~Our sensitivity study extends and refine previous works in terms of background models, statistical methods, and treatment of the underlying astrophysical uncertainties.}

\begin{document}
\maketitle
\flushbottom

\section{Introduction}
The presence of Dark Matter (DM) in the Universe has firmly been established through increasingly accurate cosmological observations~\cite{Bertone:2016nfn}.~Evidence has been gathered in a wide range of physical scales, from sub-galactic scales to the largest scales we can probe in the Universe~\cite{Bertone:2004pz}.~This includes data on the vertical motion of stars in the solar neighbourhood~\cite{Kuijken:1989hu}, the rotation curve of spiral galaxies~\cite{Rubin:1970zza}, the velocity dispersion of galaxies in galaxy clusters~\cite{Zwicky}, gravitational lensing events~\cite{Kaiser:1992ps}, the dynamics of colliding clusters~\cite{Clowe:2006eq}, the large-scale cosmological structures~\cite{Blumenthal:1984bp} and anisotropies in the cosmic microwave background temperature~\cite{Ade:2015xua}.~While the evidence for DM is strong, it is entirely based on gravitational effects, directly or indirectly related to the gravitational pull that DM exerts on visible matter and light.~As a result, we still do not know whether or not DM is made of particles which have so far escaped detection.~One promising approach to answer this question is the so-called DM direct detection technique~\cite{Drukier:1983gj,Goodman:1984dc}.

Direct detection experiments search for DM-nucleus or -electron scattering events in low-background detectors located deep underground~\cite{Undagoitia:2015gya}.~Next generation direct detection experiments searching for signals of DM-electron interactions with germanium and silicon semiconductor detectors of mass in the 0.1 - 1 kg range~\cite{Agnese:2016cpb,Crisler:2018gci,Aguilar-Arevalo:2019wdi} are of special interest to this work.~In these detectors, the energy deposited in a DM-electron scattering event can cause an observable electronic transition from the valence to the conduction band of the semiconductor target.~For kinematical reasons, this detection principle can outperform methods based on nuclear recoils in the search for DM particles in the 1 MeV to 1 GeV mass range~\cite{Essig:2011nj}.~Sub-GeV DM can also be searched for with, e.g.~dual-phase argon~\cite{Agnes:2018oej} and xenon~\cite{Essig:2012yx,Essig:2017kqs,Aprile:2019xxb} targets, graphene~\cite{Hochberg:2016ntt,Geilhufe:2018gry}, 3D Dirac materials~\cite{Hochberg:2017wce,Geilhufe:2019ndy,Coskuner:2019odd}, polar crystals~\cite{Knapen:2017ekk}, scintillators~\cite{Derenzo:2016fse,Blanco:2019lrf} and superconductors \cite{Hochberg:2015pha,Hochberg:2015fth,Hochberg:2019cyy}.~For a comparison of the performance of different materials in the search for sub-GeV DM, see~\cite{Griffin:2019mvc}.

For the purposes of this paper, we divide detectors based on germanium and silicon semiconductor crystals into two categories:~1) ``high-voltage'' (HV) detectors; and 2) ``charge-coupled device'' (CCD) detectors.~Detectors operating in HV mode have the capability to amplify the small charge produced by DM scattering in target crystals into a large phonon signal by applying a bias of about 100~V across the detector and exploiting the so-called Neganov-Trofimov-Luke effect~\cite{Luke1988Dec}.~The SuperCDMS experiment demonstrated that this approach allows to achieve sensitivity to single-charge production~\cite{Agnese:2018col}.~Similarly, CCD sensors can achieve single-charge sensitivity by measuring the charge collected by single pixels in the CCD device exploiting ultra-low readout noise techniques, as in ``Skipper'' CCDs~\cite{Abramoff:2019dfb}.~Currently operating experiments belonging to the first category include the HV mode run of the silicon SuperCDMS experiment, which delivered data corresponding to an exposure of 0.49 gram-days~\cite{Agnese:2018col}.~The null result reported by this search has been used to set a 90\% C.L exclusion limit of $10^{-30}$~cm$^2$ on the cross section for DM-electron scattering for DM-particle masses around 1 MeV.~For the future, the SuperCDMS collaboration plans to operate in the HV mode larger germanium and silicon detectors, reaching an exposure of about 50 and 10 kg-year, respectively~\cite{Agnese:2016cpb}.~Operating DM direct detection experiments belonging to the second category include the DAMIC and SENSEI experiments, both using CCD silicon sensors.~For example, the null result of the SENSEI experiment has been used to set 90\% C.L exclusion limits on the DM-electron scattering cross section for DM particle masses in the 0.5 - 100 MeV range, with a minimum excluded cross section of about $5\times 10^{-35}$~cm$^2$ for a DM mass of about 10 MeV (and short-range interactions)~\cite{Abramoff:2019dfb}.~For the future, the SENSEI and DAMIC collaborations aim at building CCD silicon detectors of 0.1 kg and 1 kg target mass, respectively~\cite{Crisler:2018gci,Aguilar-Arevalo:2019wdi}.

Predictions for the rate of DM-induced electronic transitions in semiconductor crystals depend on a number of theoretical and experimental inputs~\cite{Essig:2011nj,Graham:2012su,Lee:2015qva,Essig:2015cda,Roberts:2016xfw,Crisler:2018gci,Agnese:2018col,Abramoff:2019dfb,Aguilar-Arevalo:2019wdi,Catena:2019gfa}.~Firstly, they depend on the assumed DM-electron interaction model, e.g. on whether the interaction is long- or short-range and on its Lorentz structure~\cite{Catena:2019gfa}.~Secondly, they depend on the semiconductor band structure, and in particular on the initial and final state electron energy and wave functions~\cite{Essig:2015cda}.~They also depend on astrophysical inputs, such as the local DM density and velocity distribution, as well as on detector characteristics such as energy threshold, efficiency, and energy deposition to number of produced electron-hole pairs conversion, just to name a few.

Motivated by the recent experimental results reviewed above and by the improved understanding of the experimental backgrounds that these results have produced, this article aims at assessing the sensitivity to DM particles in the sub-GeV mass range of future direct detection experiments using germanium and silicon semiconductor crystals as target materials.~We address this problem focusing on the so-called ``dark photon'' model~\cite{Holdom:1985ag,Essig:2011nj} as a framework to describe the interactions of DM in semiconductor crystals.~We compute the projected sensitivities of future germanium and silicon detectors by comparing the null, i.e.~background-only hypothesis to the alternative,~i.e.~background plus signal hypothesis using the likelihood ratio as a test statistic~\cite{Cowan:2010js}.~Doing so, we provide a detailed description of the background models used in our analysis.~We present our results in terms of the DM-electron scattering cross section required to reject the null hypothesis in favour of the alternative one with a statistical significance corresponding to 3 or 5 standard deviations.~We compute the significance for DM particle discovery using asymptotic formulae for the probability density function of the likelihood ratio~\cite{Cowan:2010js}, after explicitly validating them by means of Monte Carlo simulations.~We also test the stability of our results under variations in the underlying astrophysical inputs.~Our sensitivity study extends previous works, e.g.~\cite{Essig:2015cda}, by:~1) adopting a refined experimental background model 2) computing the projected sensitivity by using the likelihood ratio method; and 3) exploring the dependence of our results on the DM space and velocity distribution.

This paper is organised as follows.~In Sec.~\ref{sec:theory}, we review the theory of DM-electron scattering in silicon and germanium semiconductor crystals, while in Sec.~\ref{sec:detectors} we describe the efficiency, energy threshold and experimental backgrounds assumed when modelling future HV germanium and silicon detectors, as well as future silicon CCD detectors.~We present our methodology and projected sensitivities to sub-GeV DM particles in Sec.~\ref{sec:results} and conclude in Sec.~\ref{sec:conclusions}.

\section{Dark matter scattering in semiconductor crystals}
\label{sec:theory}
In this section, we review the theory of DM-electron scattering in semiconductor crystals. We start by presenting a general expression for the rate of DM-induced electronic transitions in condensed matter systems (Sec.~\ref{sec:crystals}).~We then specialise this expression to the case of electronic transitions from the valence to the conduction band of germanium and silicon semiconductors (Sec.~\ref{sec:band}).~As we will see, the transition rate found in Sec.~\ref{sec:band} depends on an integral over DM particle velocities (Sec.~\ref{sec:kin}) and on the amplitude for DM scattering by free electrons (Sec.~\ref{sec:free}).

\subsection{Dark matter-induced electronic transitions}
\label{sec:crystals}
The rate of DM-induced transitions from an initial electron state $|\mathbf{e}_1\rangle$ to a final electron state $|\mathbf{e}_2\rangle$ is~\cite{Catena:2019gfa}
\begin{align}
\mathscr{R}_{1\rightarrow 2}&=\frac{n_{\chi}}{16 m^2_{\chi} m^2_e}
\int \frac{{\rm d}^3 q}{(2 \pi)^3} \int {\rm d}^3 v f_{\chi}(\mathbf{v}) (2\pi) \delta(E_f-E_i) \overline{\left| \mathcal{M}_{1\rightarrow 2}\right|^2}\, ,\label{eq:transition rate}
\end{align}
where $m_\chi$ is the DM particle mass, while $n_\chi=\rho_\chi/m_\chi$ and $f_\chi(\mathbf{v})$ are the local DM number density and velocity distribution, respectively.~If not otherwise specified, we set the local DM mass density $\rho_\chi$ to $0.4$~GeV/cm$^3$~\cite{Catena:2009mf}.~In all applications, for the local DM velocity distribution we assume~\cite{Lewin:1995rx}
\begin{align}
    f_\chi(\mathbf{v})&= \frac{1}{N_{\rm esc}\pi^{3/2}v_0^3}\exp\left[-\frac{(\mathbf{v}+\mathbf{v}_\oplus)^2}{v_0^2} \right]
    \times\Theta\left(v_{\rm esc}-|\mathbf{v}+\mathbf{v}_\oplus|\right)\, ,
    \label{eq:fv}
\end{align}
that is, a truncated Maxwell-Boltzmann~distribution boosted to the detector rest frame.~Here $N_{\rm esc}\equiv \erf(v_{\rm esc}/v_0)-2 (v_{\rm esc}/v_0)\exp(-v_{\rm esc}^2/v_0^2)/\sqrt{\pi}$ implies that $f_\chi(\mathbf{v})$ is unit-normalised.~In Sec.~\ref{sec:results}, we present our results by varying most probable speed~$v_0$, detector's velocity $v_\oplus$ and galactic escape velocity $v_{\rm esc}$ within their experimental uncertainties (see Sec.~\ref{sec:astro} for further details).~The squared electron transition amplitude, $\overline{\left| \mathcal{M}_{1\rightarrow 2}\right|^2}$, depends on the initial and final state electron wave functions, $\psi_1$ and $\psi_2$, respectively, and on the amplitude for DM scattering by free electrons, $\mathcal{M}$.~Without any further restriction on the amplitude $\mathcal{M}$, it can be written as~\cite{Catena:2019gfa}
\begin{align}
    \overline{\left| \mathcal{M}_{1\rightarrow 2}\right|^2}\equiv \overline{\left|\int  \frac{{\rm d}^3 k}{(2 \pi)^3} \, \psi_2^*(\mathbf{k}+\mathbf{q})  
\mathcal{M}(\mathbf{q},\mathbf{v}_{\rm el}^\perp)
\psi_1(\mathbf{k}) \right|^2}\, , \label{eq:transition amplitude}
\end{align}
where a bar denotes an average (sum) over initial (final) spin states.~Here, $\mathbf{q}=\mathbf{p}-\mathbf{p}'$, with $\mathbf{p}$ and $\mathbf{p}'$ initial and final DM particle momenta, respectively, is the momentum transfer and we introduced 
\begin{align}
\mathbf{v}_{\rm el}^\perp &= \frac{\left( \mathbf{p} + \mathbf{p}' \right)}{2 m_{\chi}} - \frac{\left( \mathbf{k} + \mathbf{k}' \right)}{2 m_e}  =\mathbf{v} - \frac{\mathbf{q}}{2\mu_{\chi e}} - \frac{\mathbf{k}}{m_e}\, ,
\end{align}
where $\mathbf{v}\equiv \mathbf{p}/m_\chi$ is the incoming DM particle velocity, $m_e$ the electron mass, and $\mu_{\chi e}$ the reduced DM-electron mass. If the DM-electron scattering were elastic, $\mathbf{v}_{\rm el}^\perp\cdot \mathbf{q}=0$ would apply, justifying the notation.
The initial and final state energies in Eq.~(\ref{eq:transition rate}) are defined as follows,
\begin{align}
    E_i &= m_\chi + m_e + \frac{m_\chi}{2}v^2 + E_1\, , \label{eq: energy initial}\\
    E_f &= m_\chi + m_e + \frac{|m_\chi\mathbf{v}-\mathbf{q}|^2}{2m_\chi} + E_2\,, \label{eq: energy final}
\end{align}
where we denote the electron initial and final energy by $E_1$ and $E_2$, and their difference by $\Delta E_{1\rightarrow 2}=E_2-E_1$.

In the case of the ``dark photon'' model for DM-electron interactions (introduced below in Sec.~\ref{sec:free}), the free electron scattering amplitude only depends on $q=|\mathbf{q}|$, $\mathcal{M}=\mathcal{M}(q)$, and Eq.~(\ref{eq:transition rate}) simplifies to~\cite{Catena:2019gfa}
\begin{align}
\mathscr{R}_{1\rightarrow 2}&=\frac{n_{\chi}}{16 m^2_{\chi} m^2_e}
\int \frac{{\rm d}^3 q}{(2 \pi)^3} \int {\rm d}^3 v f_{\chi}(\mathbf{v}) (2\pi) \delta(E_f-E_i) \overline{\left| \mathcal{M}(q) \right|^2} \left| f_{1\rightarrow 2}(\mathbf{q})\right|^2 \, ,
\label{eq:transition rate2}
\end{align}
where $f_{1\rightarrow 2}$ is a scalar atomic form factor measuring the initial and final state wave function overlap~\footnote{As shown in~\cite{Catena:2019gfa}, vectorial atomic form factors might also arise within a general treatment of DM-electron interactions.},
\begin{align}
    f_{1\rightarrow 2}(\mathbf{q}) \equiv \int  \frac{{\rm d}^3 k}{(2 \pi)^3} \, \psi_2^*(\mathbf{k}+\mathbf{q})  \psi_1(\mathbf{k}) \,.
\label{eq:f12}
\end{align}

\subsection{Crystal form factors}
\label{sec:band}
The above expressions refer to generic $|\mathbf{e}_1\rangle\rightarrow|\mathbf{e}_2\rangle$ electronic transitions.~We now specialise them to the case of transitions from a valence to the conduction band in a semiconductor crystal.~In the case of crystals, electron states are labelled by a band index ``$i$'' and a wavevector ``$\mathbf{k}$'' in the first Brillouin zone (BZ).~In Bloch form, the associated wave functions can be expressed as~\cite{Essig:2015cda}
\begin{align}
\psi_{i\mathbf{k}}(\mathbf{x}) = \frac{1}{\sqrt{V}} \sum_{\mathbf{G}} u_i(\mathbf{k}+\mathbf{G}) e^{i(\mathbf{k}+\mathbf{G})\cdot\mathbf{x}} \,,
\end{align}
where $V$ is the volume of the crystal, i.e. $\int d^3{x} \, e^{i\mathbf{k}\cdot \mathbf{x}} = (2\pi)^3 \delta^{(3)}(\mathbf{k})$ and $V=(2\pi)^3 \delta^{(3)}(0)$, while $\mathbf{G}$ is the reciprocal lattice vector.~For the wave functions to be unit-normalised, the $u_i$~coefficients must fulfil  
\begin{align}
\sum_{\mathbf{G}}|u_i(\mathbf{k}+\mathbf{G})|^2 = 1\,.
\end{align}
With these definitions, we can now interpret the transition rate in Eq.~(\ref{eq:transition rate}) as the transition rate, $\mathscr{R}_{i\mathbf{k}\rightarrow i'\mathbf{k}'}$, from the valence level $\{i,\mathbf{k}\}$ to the conduction level $\{i',\mathbf{k}'\}$.~Summing over all final state energy levels, and all filled initial state energy levels (while taking into account the initial state electron spin degeneracy), the resulting transition rate, $\mathscr{R}_{\rm crystal}$, reads as follows~\cite{Essig:2015cda}
\begin{align}
\label{eq:Rcrystal}
\mathscr{R}_{\rm crystal} &= 2\sum_{i}\int_{\rm BZ} \frac{V {\rm d^3} k}{(2\pi)^3}\sum_{i'}\int_{\rm BZ} \frac{V {\rm d^3} k'}{(2\pi)^3}\, \mathscr{R}_{i\mathbf{k}\rightarrow i'\mathbf{k}'} \,,\nonumber\\
&=  \frac{\rho_\chi}{m_\chi} \frac{N_{\rm cell} \alpha}{16 \pi m_\chi^2} \int {\rm d}\ln E_e \int {\rm d}\ln q \left(\frac{E_e}{q} \right) \eta(v_{\rm min}(q,E_e))\overline{\left| \mathcal{M}(q) \right|^2} \left| f_{\rm crystal}(q,E_e) \right|^2 \,,
\end{align}
where the deposited energy $E_e$ is defined as $E_e\equiv\Delta E_{1\rightarrow 2}=E_{i'\mathbf{k}'}-E_{i\mathbf{k}}$, the free electron scattering amplitude, $\mathcal{M}$, is assumed to be a function of $q$ only, as in the case of the dark photon model, and $V=N_{\rm cell} V_{\rm cell}$.~Here $V_{\rm cell}$ is the volume of individual cells and $N_{\rm cell}=M_{\rm target}/M_{\rm cell}$ is the number of cells in the crystal, where $M_{\rm cell}=2m_{\rm Ge} = 135.33$~GeV for germanium and $M_{\rm cell}=2m_{\rm Si} = 52.33$~GeV for silicon~\cite{Essig:2015cda}, while $M_{\rm target}$ is the detector target mass.~The velocity integral in Eq.~(\ref{eq:transition rate}) is now reabsorbed in the definition of the $\eta(v_{\rm min})$~function (given below in Sec.~\ref{sec:kin}), while the crystal form factor $|f_{\rm crystal}(q,E_e)|^2$ is defined as~\cite{Essig:2015cda}
\begin{align}
\left| f_{\rm crystal}(q,E_e) \right|^2 &= \frac{2 \pi^2 (\alpha m_e^2 V_{\rm cell})^{-1}}{E_e} \sum_{ii'} \int_{\rm BZ} \frac{V_{\rm cell} {\rm d}^3k}{(2\pi)^3}\int_{\rm BZ} \frac{V_{\rm cell} {\rm d}^3k'}{(2\pi)^3} \nonumber\\
&\times E_e\delta(E_e-E_{i'\mathbf{k}'}+E_{i\mathbf{k}}) \sum_{\mathbf{G}'} q\delta(q-|\mathbf{k}'-\mathbf{k}+\mathbf{G}'|) \left| f_{[i\mathbf{k},i'\mathbf{k}',\mathbf{G}]} \right|^2 \,,
\label{eq:crystalff}
\end{align}
where 
\begin{align}
f_{[i\mathbf{k},i'\mathbf{k}',\mathbf{G}]} = \sum_{\mathbf{G}} u_{i'}^*(\mathbf{k}'+\mathbf{G}+\mathbf{G}')u_i(\mathbf{k}+\mathbf{G})
\end{align}
and $\alpha$ is the fine structure constant.~In the numerical applications, we use germanium and silicon crystal form factors found in~\cite{Essig:2015cda}.~Following~\cite{Essig:2015cda}, we set $2\pi^2(\alpha m_e^2V_{\rm cell})^{-1}=1.8$~eV for germanium and $2\pi^2(\alpha m_e^2V_{\rm cell})^{-1}=2.0$~eV for silicon.~Finally, we rewrite Eq.~(\ref{eq:Rcrystal}) in differential form, obtaining the differential rate of electronic transitions in germanium and silicon crystals,
\begin{align}
\frac{{\rm d}\mathscr{R}_{\rm crystal}}{{\rm d} \ln E_e} = 
\frac{\rho_\chi}{m_\chi} \frac{N_{\rm cell} \alpha}{16 \pi m_\chi^2}  \int {\rm d}\ln q \left(\frac{E_e}{q} \right) \eta(v_{\rm min}(q,E_e))\overline{\left| \mathcal{M}(q) \right|^2} \left| f_{\rm crystal}(q,E_e) \right|^2 \,.
\label{eq:dRcrystal}
\end{align}
In order to compare Eq.~(\ref{eq:dRcrystal}) with observations, one has to convert $E_e$ into a number of electron-hole pairs produced in a DM-electron scattering event, $Q$.~The two quantities can be related as follows~\cite{Essig:2015cda}
\begin{align}
Q(E_e) = 1 + \lfloor (E_e - E_{\rm gap})/\varepsilon \rfloor\,,
\label{eq:conv}
\end{align}
where $\lfloor \cdot \rfloor$ is the floor function.~The observed band-gap, $E_{\rm gap}$, and mean energy per electron-hole pair, $\varepsilon$, are $E_{\rm gap}=0.67$~eV and $\varepsilon=2.9$~eV for germanium, while $E_{\rm gap}=1.11$~eV and $\varepsilon=3.6$~eV for silicon.

\subsection{Kinematics}
\label{sec:kin}
The $\eta(v_{\rm min})$ function in Eq.~(\ref{eq:Rcrystal}) depends on the velocity distribution $f_\chi$ via a three-dimensional integral,
\begin{align}
\eta(v_{\rm min}(q,E_e)) = \int {\rm d}^3 v \frac{f_\chi(\mathbf{v})}{v} \Theta\left(v - v_{\rm min}(q,E_e)\right)\,,
\label{eq:eta}
\end{align}
where $v=|\mathbf{v}|$, and $v_{\rm min}(q,E_e)$ is the minimum velocity required to induce a transition between two electronic states separated by the energy gap $E_e$  when the momentum transferred in the process is $q$,
\begin{align}
v_{\rm min}(q,E_e) =  \frac{E_e}{q} + \frac{q}{2 m_\chi}\,. 
\label{eq:vmin} 
\end{align}
The $\Theta$ function in Eq.~(\ref{eq:eta}) arises from the integration over the momentum transfer in Eq.~(\ref{eq:transition rate}).~The minimum velocity $v_{\rm min}(q,E_e)$ can also be derived from energy conservation, which implies 
\begin{align}
    \mathbf{v}\cdot \mathbf{q} = E_e+\frac{q^2}{2m_\chi}\, . 
    \label{eq: energy conservation v.q}
\end{align}
Maximising Eq.~(\ref{eq: energy conservation v.q}) with respect to the momentum transfer $\mathbf{q}$ for a given energy gap $E_e$ gives back Eq.~(\ref{eq:vmin}).~For the velocity distribution in Eq.~(\ref{eq:fv}) (and for $\mathcal{M}$ depending on $q$ only), the velocity integral in Eq.~(\ref{eq:dRcrystal}) can be evaluated analytically.~For the result of this integration, see e.g.~\cite{Lewin:1995rx}.

\subsection{Dark matter-electron interaction model}
\label{sec:free}
In order to evaluate Eq.~(\ref{eq:Rcrystal}), we need to specify a model for DM-electron interactions from which to calculate $\mathcal{M}$.~In this analysis, we focus on the so-called dark photon model~\cite{Holdom:1985ag,Essig:2011nj}, which arises as an extension of the Standard Model (SM) of particle physics.~In the dark photon model, the SM is extended by one DM candidate and one additional $U(1)$ gauge group under which only the DM particle candidate is charged.~The associated gauge boson, here denoted by $A'_\mu$, is the dark photon.~Radiative corrections are generically expected to generate a kinetic mixing term between dark and ordinary photon, i.e.~$\epsilon F_{\mu\nu} F^{'\mu\nu}$ where $F_{\mu\nu}$  ($F^{'\mu\nu}$) is the photon (dark photon) field strength tensor and $\epsilon$ is a dimensionless mixing parameter.~This kinetic mixing acts as a portal between the DM and SM sectors.~After a field redefinition which diagonalises the photon and dark photon kinetic terms, and assuming that the DM candidate is a Dirac fermion, the dark photon model can be formulated in terms of the following Lagrangian~\cite{Holdom:1985ag,Essig:2011nj}
\begin{align}
\mathscr{L} &= \mathscr{L}_{\rm SM} -\frac{1}{4} F^{'}_{\mu\nu} F^{'\mu\nu} + \frac{1}{2} m_{A'}^2 A_{\mu}^{'} A^{'\mu}  \nonumber\\
& + \sum_i \bar{f}_i \left( -e q_i \gamma^\mu A_\mu - \epsilon e q_i \gamma^\mu A^{'}_\mu  - m_i \right) f_i \nonumber\\
& + \bar{\chi} (-g_D\gamma^\mu A_\mu^{'} - m_\chi) \chi\,,
\label{eq:LDP}
\end{align}
where $g_D$ is the gauge coupling associated with the additional $U(1)$ group, $m_{A'}$ and $m_{\chi}$ are the dark photon and DM particle mass, respectively, while $\chi$ and $f_i$ are four-component Dirac spinors for the DM  particle and the SM fermions, respectively.~In the second line of Eq.~(\ref{eq:LDP}), we denote by $q_i$ ($m_i$) the electric charge (mass) of the $f_i$ SM fermion.~For the purposes of this analysis, we do not need to specify a mechanism for the spontaneous breaking of the additional $U(1)$ gauge group and the generation of the dark photon and DM particle mass.~Within the dark photon model the squared modulus of the amplitude for DM-electron scattering can be written as follows~\cite{Essig:2015cda}
\begin{align}
\overline{\left| \mathcal{M}(q)  \right|^2} = \frac{16\pi m_\chi^2 m_e^2}{\mu^2_{\chi e}} \sigma_e \left| F_{\rm DM} (q) \right|^2 \,,
\end{align}
where $\mu_{\chi e}$ is the reduced DM-electron mass, $\sigma_e$ is a reference scattering cross section setting the strength of DM-electron interactions and $F_{\rm DM}$ is the ``DM form factor'' which encodes the $q$-dependence of the amplitude.~It reads $F_{\rm DM}(q)=1$ for $q^2\ll m^2_{A'}$ (short-range or contact interaction) and $F_{\rm DM}(q)=q_{\rm ref}^2/q^2$ for $q^2\gg m^2_{A'}$ (long-range interaction), where we set the reference momentum $q_{\rm ref}$ to the value $q_{\rm ref}=\alpha m_e$, the typical momentum transfer in DM-induced electronic transitions.

\section{Detector models}
\label{sec:detectors}
In this section we specify the experimental inputs we use and the assumptions we make to compute the statistical significance for DM discovery at direct detection experiments using germanium and silicon semiconductor crystals as target materials.~We consider two classes of detectors separately:~germanium and silicon HV detectors, as in future runs of the SuperCDMS experiment, and silicon CCDs, as in future runs of the DAMIC and SENSEI experiments.

\subsection{High voltage Si/Ge detectors}
We refer to HV detectors as experimental devices resembling the SuperCDMS experiment in the operating mode described in the recent analysis~\cite{Agnese:2018col}.~In this configuration, SuperCDMS exploits a 0.93 g high-purity silicon crystal instrumented on one side with transition-edge sensors and on the other side with an electrode made of an aluminum-amorphous silicon bilayer.~This device can achieve single-charge resolution by exploiting the Neganov-Trofimov-Luke effect~\cite{Luke1988Dec}.~It consists in the emission of phonons generated by electron-hole pairs drifting across a bias of 140 V (the high voltage defining this operating mode).~This effect can amplify the small charge signal associated with DM scattering in a HV detector into a large phonon signal.

In this work, we investigate the sensitivity of next generation HV detectors, taking the expected reach of SuperCDMS as a guideline~\cite{Agnese:2016cpb}.~Doing so, we focus on germanium and silicon targets separately, in that different exposures are planned for the two targets.~More specifically, in the case of germanium HV detectors, we assume an exposure of 50 kg-year.~For HV detectors using silicon targets, we assume an exposure of 10 kg-year (see Tab.~4 in~\cite{Agnese:2016cpb} for further details).

Another important experimental input to our analysis is the detection efficiency of HV detectors.~In general, only a fraction of DM-induced electronic transitions is expected to be successfully recorded by detectors used in DM direct detection experiments.~The fraction of events that are successfully detected is called the detection efficiency.~For both silicon and germanium HV detectors, we assume an energy-independent (i.e. $E_e$-independent) detection efficiency of 90\%, which is expected to be a fairly good approximation for $Q$ between 1 and~8~\cite{Agnese:2018col}.~When computing the expected number of DM signal events in a given $Q$-bin, we then multiply Eq.~(\ref{eq:dRcrystal}) by a detection efficiency factor $\xi_i=0.9$ for both germanium and silicon HV detectors.~Here, ``$i$'' is an index labelling the $Q=i$ bin (see Sec.~\ref{sec:results} for further details about event binning).

We now describe the experimental background model used for HV detectors in our sensitivity study.~As demonstrated recently~\cite{Agnese:2018col}, charge leakage is the dominant background source in the search for DM-electron scattering events with HV detectors for values of Q less than 3.~Indeed, large electric fields used in HV detectors can ionise impurities within the experimental apparatus causing charge carriers to tunnel into the crystal, producing a background event.~We model the event spectrum associated with this experimental background by interpreting the events measured in~\cite{Agnese:2018col} (orange line in Fig.~3) as due to charge leakage, as the authors suggest.~For larger values of $Q$, $\beta$'s and $\gamma$'s from the decay of radioactive isotopes originating from the experimental apparatus are also important~\cite{Agnese:2016cpb}.~In our sensitivity study, we model the Compton scattering of $\gamma$-rays from the decay of heavy radioactive isotopes as described in~\cite{Barker2018Aug}.~For the deposited energy spectrum induced by Compton scattering events from radiogenic $\gamma$'s, we assume a constant function of $E_e$, i.e.~$f_C(E)=$~const., in the case of silicon HV detectors, and the following combination of error functions for germanium HV detectors~\cite{Barker2018Aug}
\begin{align}
f_{C}(E_e) = \mathscr{N}\left\{0.005 + \frac{1}{N_1} \sum_{i=K,L,M.N} 0.5 A_i \left[ 1 + \erf\left(\frac{E_e - \mu_i}{\sqrt{2}\sigma_i}\right) \right] \right\} \,,
\label{eq:fC}
\end{align}
where $\mathscr{N}$ is a normalisation constant.~In principle, Eq.~(\ref{eq:fC}) receives contributions from the $K$, $L$, $M$, and $N$ shells of germanium.~In the energy range of interest, however, only the germanium $N$ shell contribution with input parameters $A_{N}/N_1=18.70$~MeV$^{-1}$, $\mu_N=0.04$~keV, $\sigma_N=13$~eV~\cite{Barker2018Aug} needs to be considered.~Both for germanium and silicon, we conservatively normalise $f_C(E_e)$ such that when integrated over the 0 - 50~eV range it gives 0.1~counts~kg$^{-1}$day$^{-1}$~\cite{Agnese:2016cpb}.

Finally, we assume that next generation HV germanium and silicon detectors will achieve single-charge resolution, which implies a sensitivity to energy depositions as low as the crystal's band-gap, i.e.~$Q_{\rm th}=1$, where $Q_{\rm th}$ is the experimental threshold, i.e. the minimum number of detectable electron-hole pairs.

\subsection{CCD detectors}
For CCD detectors, DAMIC and SENSEI are our reference experiments.~Both experiments exploit silicon semiconductor crystals as a target and already reported results from the run of prototype detectors~\cite{Crisler:2018gci,Aguilar-Arevalo:2019wdi,Abramoff:2019dfb}.~For example, SENSEI recently reported data collected by using one silicon Skipper-CCD with a total active mass (before masking) of 0.0947 gram and consisting of $\mathcal{O}(10^6)$ pixels~\cite{Abramoff:2019dfb}.~In a CCD, DM-electron scattering events can cause electronic transitions from the valence band to the conduction band of crystals in the device's pixels.~The excited electron subsequently creates additional electron-hole pairs for each 3.6 eV of excitation energy above the band gap which are then moved pixel-by-pixel to one of the CCD corners for the read-out. 

Exploring the sensitivity to sub-GeV DM of CCD detectors, we focus on two benchmark values for the experimental exposure.~These are:~1)~100~g-year, which is the exposure SENSEI aims at~\cite{Crisler:2018gci}; and~2) 1~kg-year, as expected for the next version of the DAMIC experiment, DAMIC-M~\cite{Aguilar-Arevalo:2019wdi}.~As far as the detection efficiency of CCD detectors is concerned, we use the values reported in Tab.~I of~\cite{Abramoff:2019dfb} in the ``DM in single pixel'' line.~These values are:~$\xi_1=1$, $\xi_2=0.62$, $\xi_3=0.48$, $\xi_4=0.41$, $\xi_5=0.36$ for the $Q=1$, $Q=2$, $Q=3$, $Q=4$, and $Q=5$ bins, respectively.~We set $\xi_i=0.36$, for larger values of $Q$ ($i>5$).~Consequently, when computing the expected number of DM signal events  in the $Q=i$ bin, we multiply Eq.~(\ref{eq:dRcrystal}) by $\xi_i$.

``Dark current'' events are expected to be the dominant experimental background source for next generation CCD detectors~\cite{Abramoff:2019dfb}.~These events are evenly distributed across CCDs and are due to thermal fluctuations that excite electrons from the valence to the conduction band in crystals.~As in Tab.~1 of~\cite{Essig:2015cda}, we assume that the number, $\mathcal{N}_{i}$, of dark current events generating at least $i$ electron-hole pairs in a crystal can be estimated in terms of Poisson probabilities $\mathscr{P}$,
\begin{align}
\mathcal{N}_{i} = (n_{\rm ccd} M_{\rm pix}) (\Delta_T/{\rm hr}) \sum_{k\geq i}\mathscr{P}(\Gamma\times{\rm (counts/pixel/hr})^{-1}|k) \,,
\end{align}
where $n_{\rm ccd}$ is the number of CCDs in the detector, $M_{\rm pix}=8\times 10^6$ is the number of pixels in a CCD, $\Delta_T$ is the time of data taking in hours and $\Gamma$ is the dark current rate in counts/pixel/hr.~The number of these background events in the $Q=i$ bin is then given by $\mathcal{N}_{i}-\mathcal{N}_{i+1}$.~As anticipated, we consider two benchmark cases:~1) $n_{\rm ccd}=40$, $\Delta T=24\times365$~hr, corresponding to an exposure of 100~g-year, assuming that the mass of a single CCD is $m_{\rm ccd}=$2.5~g; and 2)~$n_{\rm ccd}=40$, $\Delta T=24\times365\times10$~hr, corresponding to an exposure of 1~kg-year, again with $m_{\rm ccd}=$2.5~g.~In each of the two scenarios  above, we present our results for two extreme values of the dark current rate, namely $\Gamma=5\times10^{-3}$~counts/pixel/day and $\Gamma=1\times10^{-7}$~counts/pixel/day.

Similarly to the case of HV silicon detectors, for the deposited energy spectrum induced by Compton scattering events from radiogenic $\gamma$'s, $f_C(E_e)$, we assume a constant function of $E_e$.~Conservatively, we normalise $f_C(E_e)$ to 0.1~counts~kg$^{-1}$day$^{-1}$ over the energy range 0 - 50~eV~\cite{Agnese:2016cpb}.

Background events due to voltage variations in the amplifiers used during read-out are assumed to be negligible, as they can be vetoed by means of a periodic read-out~\cite{Abramoff:2019dfb}.~The measured rate at SENSEI for this class of background events is of the order of $10^{-3}$~events/pixel/day~and we assume that comparable rates will be achieved at next generation silicon CCD detectors. 

Also in the case of silicon CCD detectors, we assume single-charge resolution, which implies $Q_{\rm th}=1$.

\section{Projected sensitivity}
\label{sec:results}
In Sec.~\ref{sec:stat}, we introduce the statistical method used to compute the significance for a DM particle discovery at future germanium and silicon detectors.~We present our results in Sec.~\ref{sec:num}, and investigate their stability under variations in the underlying astrophysical parameters in Sec.~\ref{sec:astro}.

\subsection{Methodology}
\label{sec:stat}
We compute the significance for DM particle discovery at a given experiment, $\mathcal{Z}$, using the likelihood ratio~\cite{Cowan:2010js},  
\begin{equation}
q_0 = -2 \ln \frac{\mathscr{L}(\mathbf{d}|0)}{\mathscr{L}(\mathbf{d}|\hat{\sigma}_e)} \,,
\label{eq:q}
\end{equation}
as a test statistic.~In Eq.~(\ref{eq:q}), $\mathscr{L}(\mathbf{d}|\sigma_e)$ is the likelihood function, $\hat{\sigma}_e$ the value of $\sigma_e$ that maximises $\mathcal{L}(\mathbf{d}|\sigma_e)$ and ${\mathbf{d}}=(\mathscr{N}_1,\dots,\mathscr{N}_n$) a dataset.~Here, $\mathscr{N}_i$ is the observed number of electronic transitions in the $i$-th $Q$-bin and the total number of bins in the $Q$ variable is assumed to be $n$.~Notice that the larger $q_0$, the worse $\sigma_e=0$ fits the data $\mathbf{d}$ and that for $\hat{\sigma}_e=0$, $q_0$ takes its minimum value, i.e. zero.~By repeatedly simulating $\mathbf{d}$ under the null hypothesis, i.e.~$\sigma_e=0$ (for given $m_\chi$, $\rho_\chi$, $v_0$, $v_{\oplus}$ and $v_{\rm esc}$), we obtain a probability density function for $q_0$ denoted here by $f_0$.~Similarly, by repeatedly simulating $\mathbf{d}$ under the alternative hypothesis, i.e. $\sigma_e=\bar{\sigma}_e\neq0$ (for given $m_\chi$, $\rho_\chi$, $v_0$, $v_{\oplus}$ and $v_{\rm esc}$), we obtain $f$, i.e.~the probability density function of $q_0$ under the alternative hypothesis.~The significance $\mathcal{Z}$ is then given by
\begin{equation}
\mathcal{Z} = \Phi^{-1}(1-p)\,,
\label{eq:Z}
\end{equation}
where $\Phi$ is the cumulative distribution function of a Gaussian probability density of mean 0 and variance 1, while
\begin{equation}
p = \int^{\infty}_{q_{\rm med}} {\rm d}q_0\,f_0(q_0)\,,
\end{equation}
and $q_{\rm med}$ is the median of $f$.~Notice that the significance given in Eq.~(\ref{eq:Z}) depends on $\bar{\sigma}_e$ as well as on the assumed values for $m_\chi$, $\rho_\chi$, $v_0$, $v_{\oplus}$ and $v_{\rm esc}$.~While both $f_0$ and $f$ can in principle be obtained via Monte Carlo simulations, asymptotically (i.e.~in the large sample limit) $f_0$ is expected to obey a ``half chi-square distribution'' for one degree of freedom, $\frac{1}{2}\chi_1^2$~\cite{Cowan:2010js}.~For a few benchmark values of $m_\chi$, $\rho_\chi$, $v_0$, $v_{\oplus}$ and $v_{\rm esc}$, we verified that $f_0$ is very well approximate by $\frac{1}{2}\chi_1^2$ by comparing the latter with the distribution of $q$ found from 10 million Monte Carlo simulations of $\mathbf{d}$.~Specifically, we find that the relative difference between the cumulative distribution functions of (the Monte Carlo generated) $f_0$ and $\frac{1}{2}\chi_1^2$ is of the order of $10^{-7}$ around $\mathcal{Z}=5$.~In order to speed up our numerical calculations, we therefore assume that the probability density function $f_0$ can be approximated by $\frac{1}{2}\chi_1^2$.~At the same time, we compute the probability density function $f$ and $q_{\rm med}$ from Monte Carlo simulations of $\mathbf{d}$.~For the likelihood, we assume 
\begin{equation}
\mathscr{L}(\mathbf{d}|\sigma_e) = \prod_{i=1}^n \frac{\left(\mathscr{B}_i+\mathscr{S}_i(\sigma_e)\right)^{\mathscr{N}_i}}{\mathscr{N}_i!} e^{-\left(\mathscr{B}_i+\mathscr{S}_i(\sigma_e)\right)} \,,
\end{equation}
where 
\begin{equation}
\mathscr{S}_i(\sigma_e) = \mathcal{E} \xi_i  \int_{Q=i}^{Q=i+1} {\rm d} Q \, \frac{{\rm d}\mathscr{R}_{\rm crystal}}{{\rm d} Q} \,,
\label{eq:S_i}
\end{equation}
while $\mathcal{E}$ is the experimental exposure and, finally, $\mathscr{B}_i$ is the total number of expected background events in the $Q=i$ bin.~In order to evaluate Eq.~(\ref{eq:S_i}), we compute ${\rm d} Q/{\rm d} E_e$ from Eq.~(\ref{eq:conv}). We introduced our assumptions for $\mathcal{E}$, $\xi_i$ and $\mathscr{B}_i$ in Sec.~\ref{sec:detectors} focusing on CCD and HV detectors separately.~When computing the joint significance for DM discovery at two experiments $A$ and $B$, we repeat the above procedure now with a likelihood function given by $\mathscr{L} = \mathscr{L}_A \mathscr{L}_B$, where $\mathscr{L}_A$ and $\mathscr{L}_B$ are the likelihood functions for the experiments $A$ and $B$, respectively.\begin{figure}[t]
\begin{center}
\begin{minipage}[t]{0.49\linewidth}
\centering
\includegraphics[width=\textwidth]{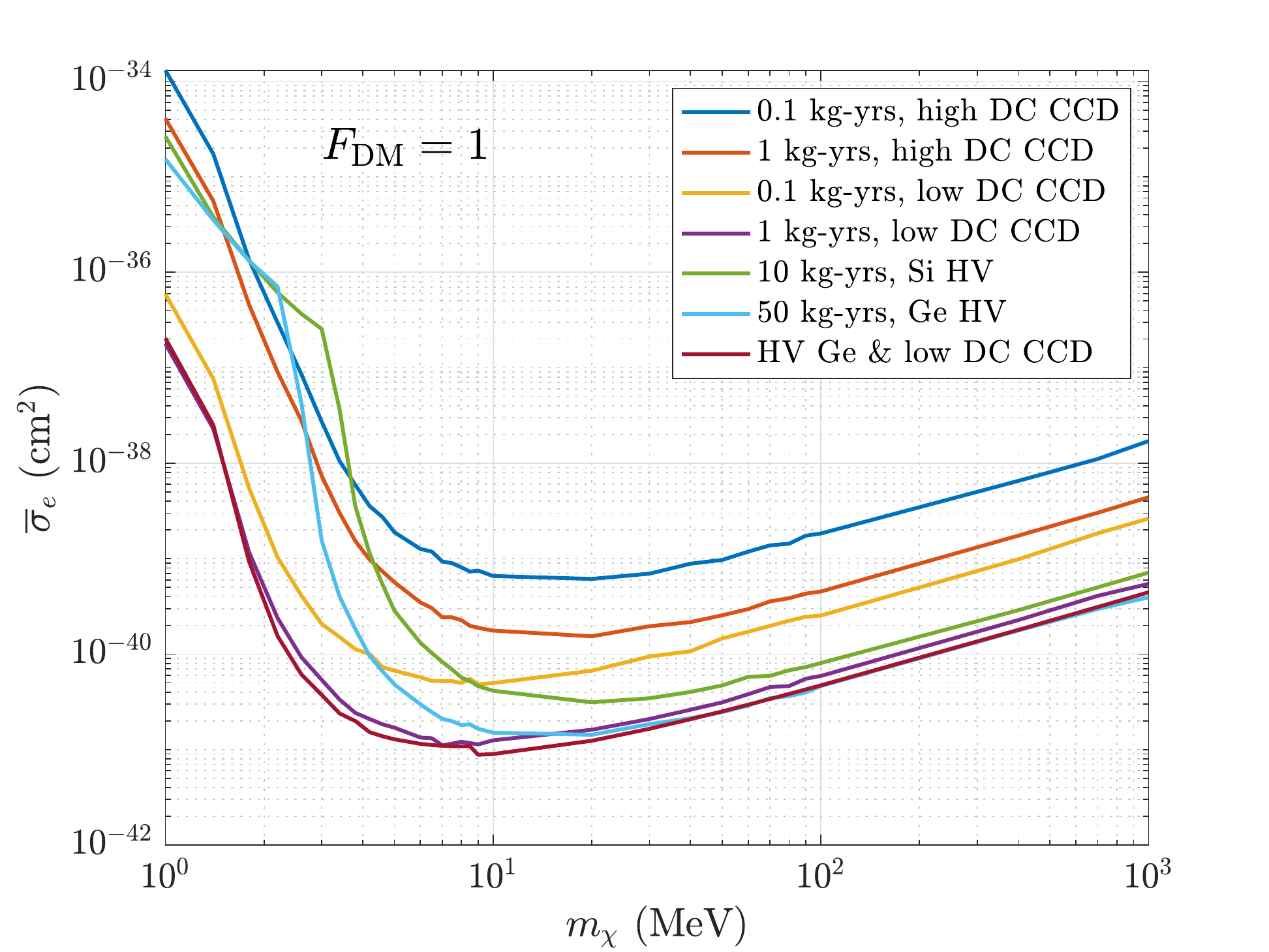}
\end{minipage}
\begin{minipage}[t]{0.49\linewidth}
\centering
\includegraphics[width=0.98\textwidth]{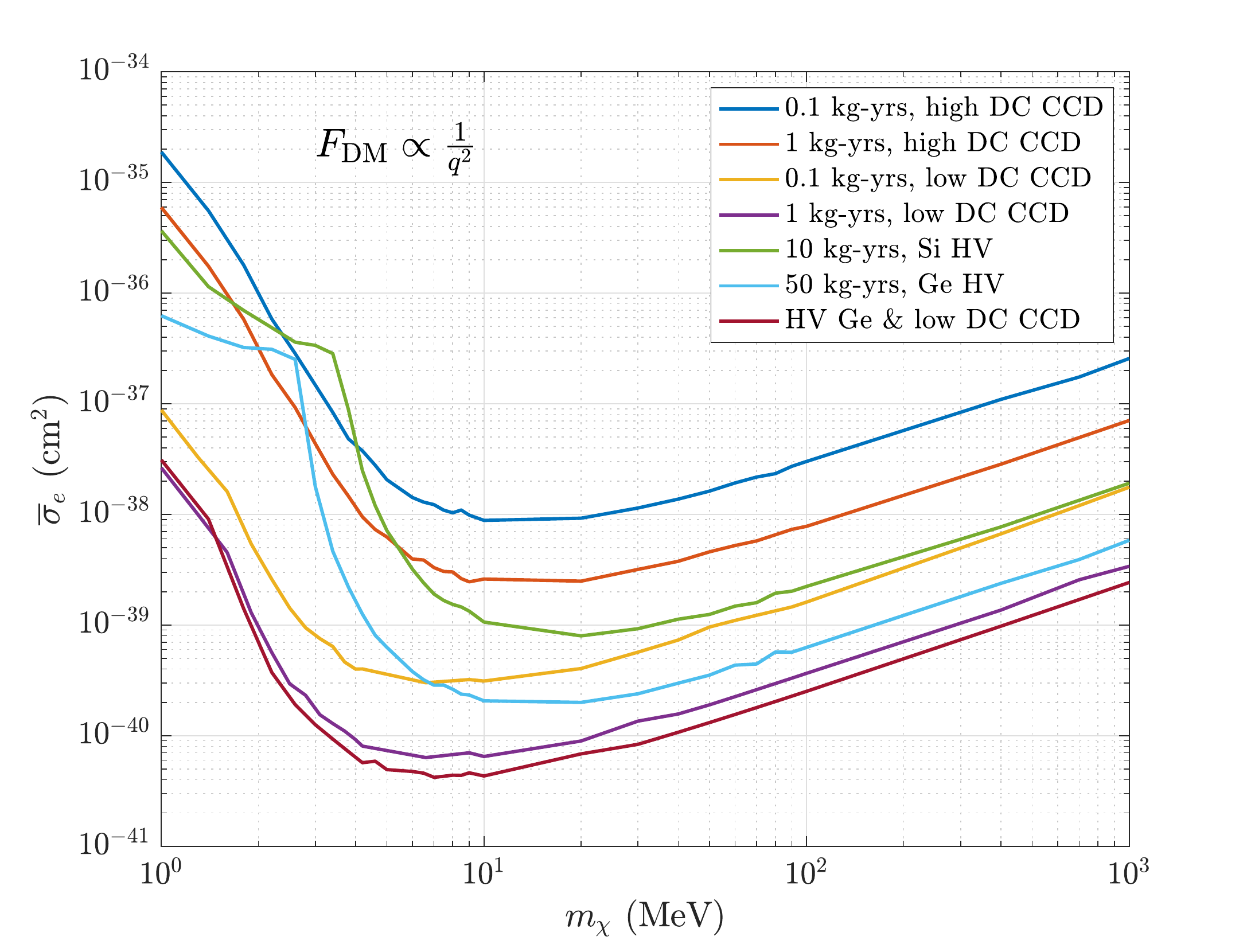}
\end{minipage}
\end{center}
\caption{Contours of constant statistical significance, $\mathcal{Z}=3$, in the ($m_\chi$, $\bar{\sigma}_e$) plane for $\rho_\chi=0.4$~GeV~cm$^{-3}$, $v_0=230$~km~s$^{-1}$, $v_{\oplus}=240$~km~s$^{-1}$ and $v_{\rm esc}=600$~km~s$^{-1}$.~Left and right panels correspond to models where $F_{\rm DM}(q^2)=1$ and $F_{\rm DM}(q^2)=q_{\rm ref}^2/q^2$, respectively.~In both panels, distinct coloured lines refer to different experimental setups:~1) CCD silicon detector with $\mathcal{E}=0.1$~kg-year and a high dark current (DC) rate, $\Gamma=5\times10^{-3}$~counts/pixel/day (blue);~2) CCD silicon detector with $\mathcal{E}=1$~kg-year and $\Gamma=5\times10^{-3}$~counts/pixel/day (red);~3) CCD silicon detector with $\mathcal{E}=0.1$~kg-year and a low DC rate of $\Gamma=1\times10^{-7}$~counts/pixel/day (orange);~4) CCD silicon detector with $\mathcal{E}=1$~kg-year and $\Gamma=1\times10^{-7}$~counts/pixel/day (violet);~5) HV silicon detector with $\mathcal{E}=10$~kg-year (green);~6) HV germanium detector with exposure of $\mathcal{E}=50$~kg-year (light blue);~and, finally 7)~a HV germanium detector with $\mathcal{E}=50$~kg-year that has reported data together with a CCD silicon detector with $\mathcal{E}=1$~kg-year and $\Gamma=1\times10^{-7}$~counts/pixel/day (brown).~Along these contours, the null hypothesis can be rejected with a significance of 3 standard deviations by one or a combinations of experiments.}  
\label{fig:Z3}
\end{figure}

\subsection{Numerical results}
\label{sec:num}
We now present the results of our sensitivity study for future DM experiments based on germanium and silicon semiconductor detectors.~We focus on the HV and CCD operating modes described in Sec.~\ref{sec:detectors} and the dark photon model reviewed in Sec.~\ref{sec:theory}.~We present our results in terms of DM-electron scattering cross section, $\bar{\sigma}_e$, required to reject the null, i.e. background only hypothesis with a statistical significance corresponding to 3 or 5 standard deviations as a function of the DM particle mass, and for benchmark values of $\rho_\chi$, $v_0$, $v_{\oplus}$ and $v_{\rm esc}$.~We investigate the dependence of our results on the local DM density and escape velocity, most probable DM speed and detector speed in the galactic rest frame in the next subsection.

Fig.~\ref{fig:Z3} shows the smallest cross section value, $\bar{\sigma}_e$, required to reject the background only hypothesis with a statistical significance of at least 3 standard deviations when the DM particle mass varies in the 1~MeV - 1~GeV range.~We obtain such $\mathcal{Z}=3$ contours using the likelihood ratio method described in Sec.~\ref{sec:stat}.~Results are presented for $\rho_\chi=0.4$~GeV~cm$^{-3}$, $v_0=230$~km~s$^{-1}$, $v_{\oplus}=240$~km~s$^{-1}$ and $v_{\rm esc}=600$~km~s$^{-1}$.~The left panel corresponds to the case $F_{\rm DM}(q^2)=1$, whereas in the right panel we assume  $F_{\rm DM}(q^2)=q_{\rm ref}^2/q^2$.~In both panels, lines with different colours correspond to distinct detectors, background assumptions or exposures.~Specifically, we consider seven different experimental setups:
\begin{enumerate}
\item CCD silicon detector operating with an exposure of $\mathcal{E}=0.1$~kg-year and a high dark current rate of $\Gamma=5\times10^{-3}$~counts/pixel/day,
\item CCD silicon detector operating with an exposure of $\mathcal{E}=1$~kg-year and a high dark current rate of $\Gamma=5\times10^{-3}$~counts/pixel/day,
\item CCD silicon detector operating with an exposure of $\mathcal{E}=0.1$~kg-year and a low dark current rate of $\Gamma=1\times10^{-7}$~counts/pixel/day,
\item CCD silicon detector operating with an exposure of $\mathcal{E}=1$~kg-year and a low dark current rate of $\Gamma=1\times10^{-7}$~counts/pixel/day,
\item HV silicon detector operating with an exposure of $\mathcal{E}=10$~kg-year,
\item HV germanium detector operating with an exposure of $\mathcal{E}=50$~kg-year, and finally
\item the case in which two distinct experiments have simultaneously reported data, namely a HV germanium detector with $\mathcal{E}=50$~kg-year and a CCD silicon detector with $\mathcal{E}=1$~kg-year and $\Gamma=1\times10^{-7}$~counts/pixel/day (i.e. the best-case scenario).
\end{enumerate}

\begin{figure}[t]
\begin{center}
\begin{minipage}[t]{0.49\linewidth}
\centering
\includegraphics[width=\textwidth]{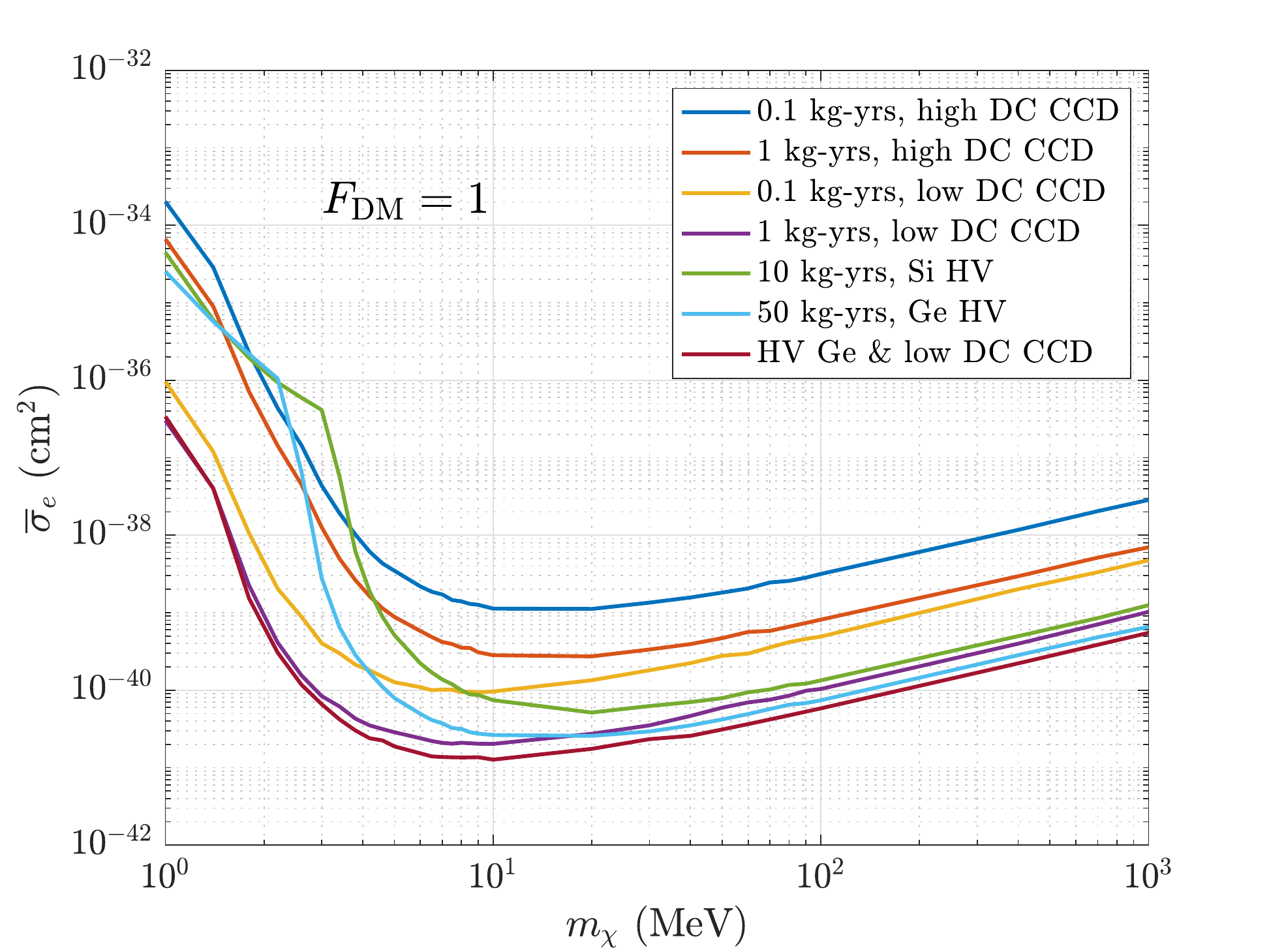}
\end{minipage}
\begin{minipage}[t]{0.49\linewidth}
\centering
\includegraphics[width=0.98\textwidth]{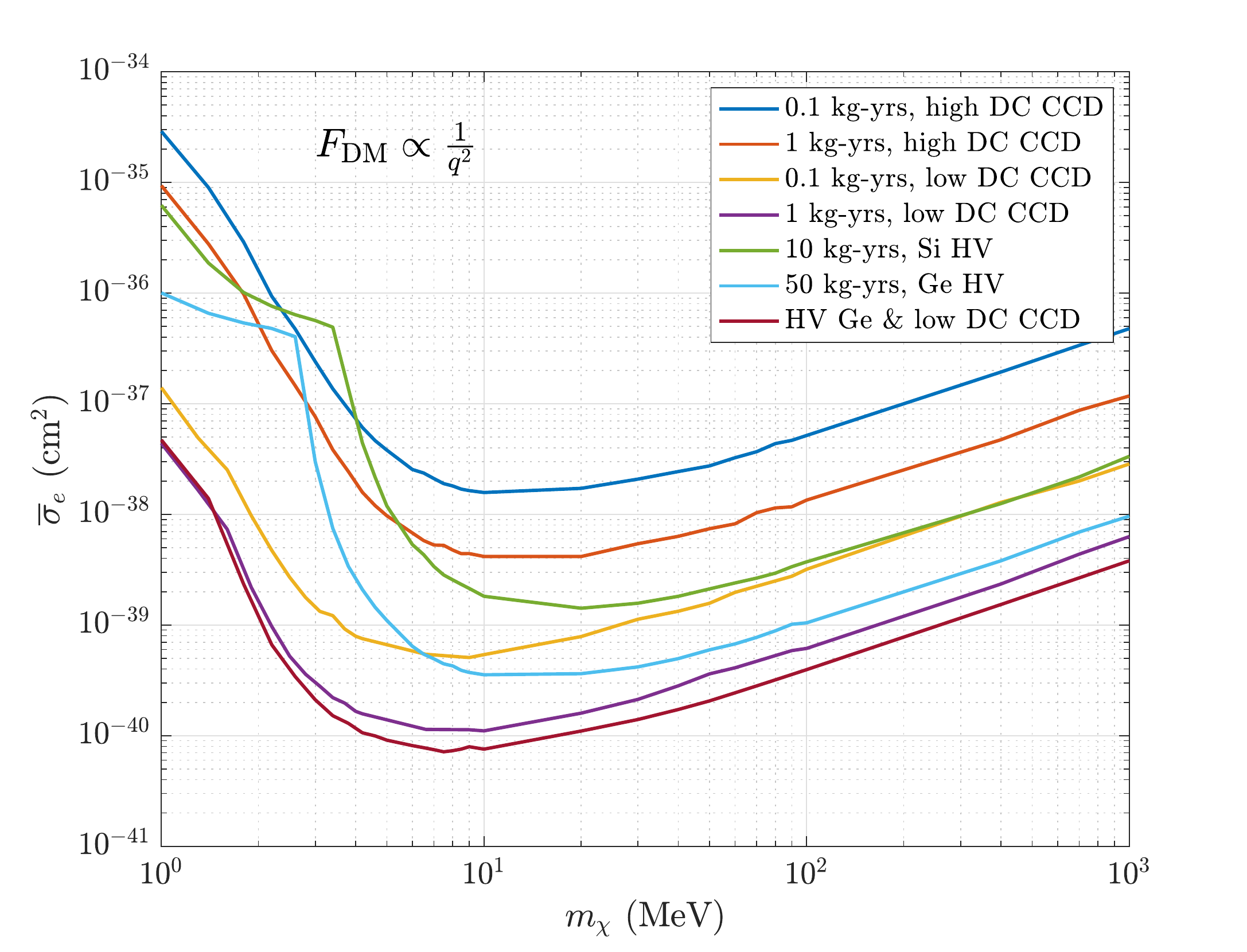}
\end{minipage}
\end{center}
\caption{Same as Fig.~\ref{fig:Z3} but for a statistical significance corresponding to 5 standard deviations, i.e.~$\mathcal{Z}=5$.}
\label{fig:Z5}
\end{figure}

For $m_\chi$ larger than about 10~MeV, HV detectors are more sensitive to DM than CCD detectors operating in the first three modes described above.~In the same mass range, however, the projected sensitivity of CCD silicon detectors operating with an exposure of $\mathcal{E}=1$~kg-year and a low dark current rate of $\Gamma=1\times10^{-7}$~counts/pixel/day is comparable with the one of a HV germanium detector operating with an exposure of $\mathcal{E}=50$~kg-year.~Below $m_\chi=10$~MeV, HV detectors rapidly lose sensitivity because of the large number and the energy spectrum of charge leakage background events assumed for HV experiments in this study (see Sec.~\ref{sec:detectors}).~In this second mass range, CCD silicon detectors are found to be more sensitive to DM than germanium and silicon HV detectors (at least for $m_\chi$ as low as 2-3 MeV).~Overall, the projected sensitivity of CCD silicon detectors operating with an exposure of $\mathcal{E}=1$~kg-year and a low dark current rate of $\Gamma=1\times10^{-7}$~counts/pixel/day is high (i.e.~the cross section corresponding to $\mathcal{Z}=3$ is comparably small) over the whole range of DM particle masses considered here.~This is especially true when $F_{\rm DM}(q^2)=q_{\rm ref}^2/q^2$, as illustrated in the right panel of Fig.~\ref{fig:Z3}.~Finally, the $\mathcal{Z}=3$ contour that we find when a HV germanium detector with $\mathcal{E}=50$~kg-year and a CCD silicon detector with $\mathcal{E}=1$~kg-year and $\Gamma=1\times10^{-7}$~counts/pixel/day have simultaneously reported a DM signal is comparable with the contour we obtain for a CCD silicon detector with $\mathcal{E}=1$~kg-year and $\Gamma=1\times10^{-7}$ (for both choices of $F_{\rm DM}$).

\begin{figure}[t]
\begin{center}
\includegraphics[width=0.5\textwidth]{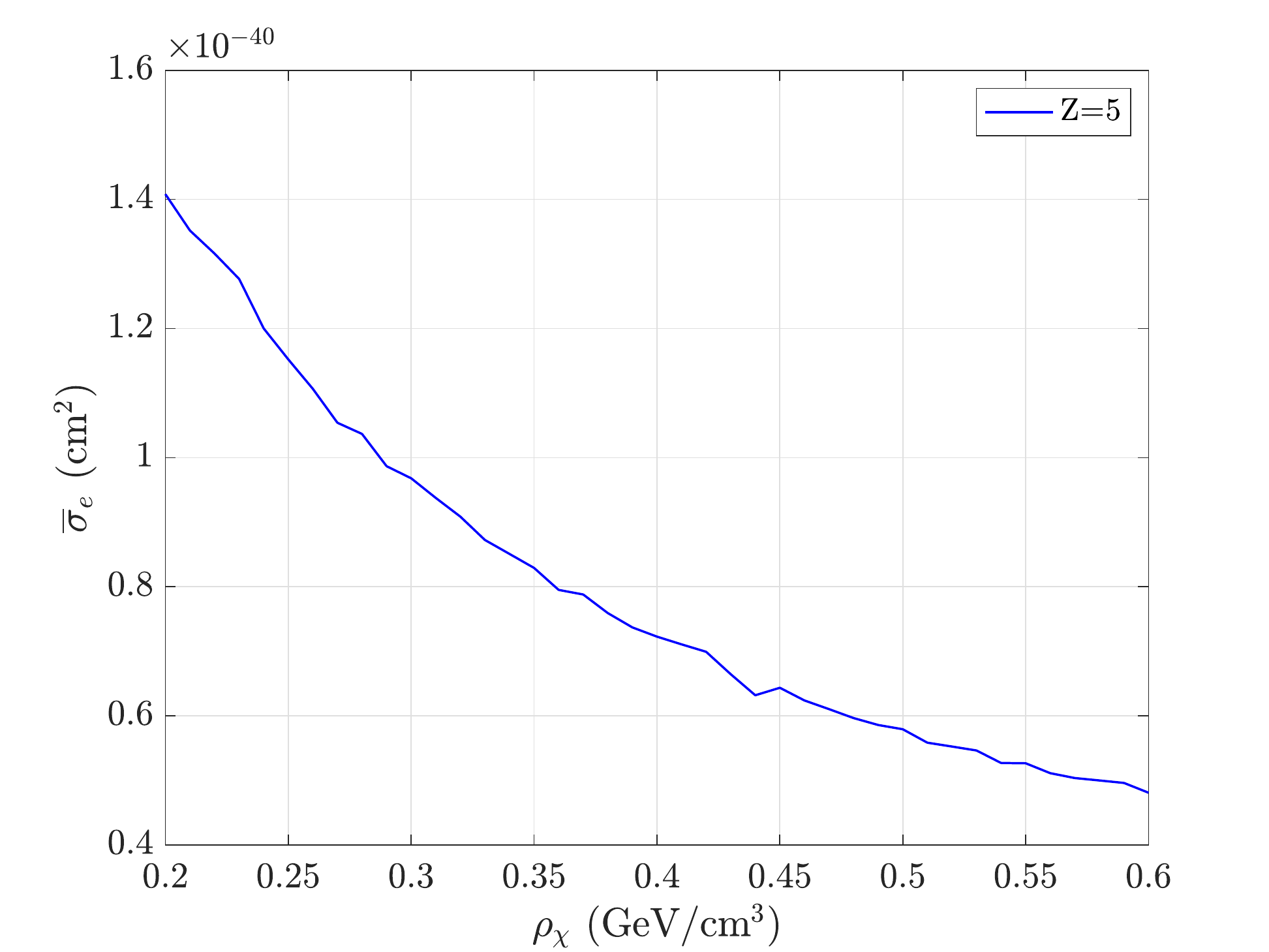}
\end{center}
\caption{DM-electron scattering cross section, $\bar{\sigma}_e$, required to reject the null hypothesis with a statistical significance corresponding to $\mathcal{Z}=5$ as a function of $\rho_\chi$ for a HV germanium detector with $\mathcal{E}=50$~kg-year, $F_{\rm DM}(q^2)=1$, $m_\chi=100$~MeV, $v_0=230$~km~s$^{-1}$, $v_{\oplus}=240$~km~s$^{-1}$ and, finally, $v_{\rm esc}=600$~km~s$^{-1}$.}  
\label{fig:rho}
\end{figure}

Similarly, Fig.~\ref{fig:Z5} shows the smallest cross section value required to exclude the background only hypothesis with a statistical significance corresponding to at least 5 standard deviations.~We obtain such $\mathcal{Z}=5$ contours in the ($m_\chi$, $\bar{\sigma}_e$) plane for $\rho_\chi=0.4$~GeV~cm$^{-3}$, $v_0=230$~km~s$^{-1}$, $v_{\oplus}=240$~km~s$^{-1}$ and $v_{\rm esc}=600$~km~s$^{-1}$.~Compared to the $\mathcal
{Z}=3$ case, the required cross section values are larger at each DM particle mass, but, as expected, above and below $m_\chi=10$~MeV the relative sensitivity of HV and CCD detectors is qualitatively unchanged.

\subsection{Astrophysical uncertainties}
\label{sec:astro}
We conclude this section by investigating the stability of our conclusions under variations of the astrophysical parameters $\rho_\chi$, $v_0$, $v_{\oplus}$ and $v_{\rm esc}$ governing the local space and velocity distribution of DM particles.~We start by focusing on the local DM density, $\rho_\chi$, upon which the rate of DM-induced electron transitions in semiconductor crystals, Eq.~(\ref{eq:dRcrystal}), linearly depends.

Fig.~\ref{fig:rho} shows the value of the DM-electron scattering cross section, $\bar{\sigma}_e$, required to reject the null hypothesis with a statistical significance corresponding to $\mathcal{Z}=5$ as a function of the local DM density for a HV germanium detector with $\mathcal{E}=50$~kg-year.~Here, we assume $F_{\rm DM}(q^2)=1$ and set $m_\chi=100$~MeV, $v_0=230$~km~s$^{-1}$, $v_{\oplus}=240$~km~s$^{-1}$ and $v_{\rm esc}=600$~km~s$^{-1}$.~As expected, we find that the value of $\bar{\sigma}_e$ solving $\mathcal{Z}=5$ is inversely proportional to $\rho_\chi$.

\begin{figure}[t]
\begin{center}
\includegraphics[width=\textwidth]{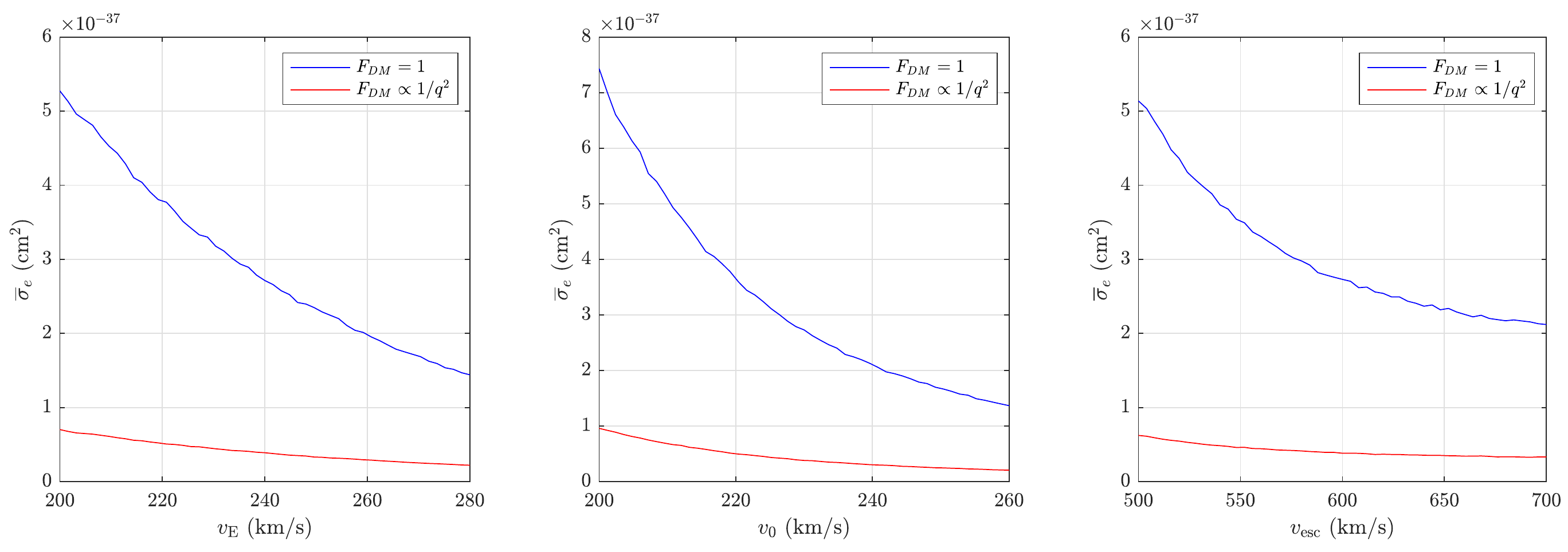}
\end{center}
\caption{Scattering cross section, $\bar{\sigma}_e$, required to reject the null hypothesis with a statistical significance of $\mathcal{Z}=5$ as a function of $v_E\equiv v_\oplus$ (left panel), $v_0$ (central panel) and $v_{\rm esc}$ (right panel) for a CCD silicon detector with $\mathcal{E}=1$~kg-year and $\Gamma=1\times10^{-7}$~counts/pixel/day.~In all panels, we consider both $F_{\rm DM}(q^2)=1$ (blue lines) and $F_{\rm DM}(q^2)=q^2_{\rm ref}/q^2$ (red lines) and set $m_\chi$ to 1~MeV and $\rho_\chi$ to $0.4$~GeV~cm$^{-3}$.~When not otherwise specified, we set  $v_0=230$~km~s$^{-1}$, $v_{\oplus}=240$~km~s$^{-1}$ and $v_{\rm esc}=600$~km~s$^{-1}$.}  
\label{fig:vel}
\end{figure}

Similarly, Fig.~\ref{fig:vel} shows the value of $\bar{\sigma}_e$ required to reject the null hypothesis with a statistical significance of $5$ as a function of the detector speed in the galactic rest frame (left panel) of the most probable DM speed (central panel) and of the local escape velocity (right panel) for a CCD silicon detector with $\mathcal{E}=1$~kg-year and $\Gamma=1\times10^{-7}$~counts/pixel/day.~In all panels, we separately consider both $F_{\rm DM}(q^2)=1$ (blue lines) and $F_{\rm DM}(q^2)=q^2_{\rm ref}/q^2$ (red lines).~Finally, we assume $m_\chi=1$~MeV and set $\rho_\chi$ to $0.4$~GeV~cm$^{-3}$.~In all panels in Fig.~\ref{fig:vel}, we find that the value of $\bar{\sigma}_e$ solving $\mathcal{Z}=5$ varies by a factor of a few over the range of astrophysical parameters considered here.~As expected, we also find that astrophysical uncertainties have a smaller impact on our results for larger values of the DM particle mass.~Notice also that in this analysis we treated $v_0$ and $v_\oplus$ as independent parameters.~This approach can account for the outcome of hydrodynamical simulations~\cite{Bozorgnia:2017brl} and generalises the so-called Standard Halo Model, where the most probable speed, $v_0$, is set to the speed of the local standard of rest~\cite{Freese:2012xd}.

\section{Conclusions}
\label{sec:conclusions}

We computed the sensitivity to DM particles in the sub-GeV mass range of future direct detection experiments using germanium and silicon semiconductor detectors.~We addressed this problem within the dark photon model for DM-electron interactions in semiconductor crystals and computed the projected sensitivities of future germanium and silicon detectors by using the likelihood ratio as a test statistic and Monte Carlo simulations.~We placed special emphasis on describing the background models used in our analysis and presented our results in terms of DM-electron scattering cross section values required to reject the background only hypothesis in favour of the background plus signal hypothesis with a statistical significance corresponding to 3 or 5 standard deviations.~We also tested the stability of our results under variations in the astrophysical parameters that govern the space and velocity distribution of DM in our galaxy.~Our sensitivity study extended previous works in terms of background models, statistical methods used to compute the projected sensitivities, and treatment of the underlying astrophysical uncertainties.~This work is motivated by the recent experimental progress, and by the improved understanding of the experimental backgrounds that this progress produced.

For $m_\chi$ larger than about 10~MeV, HV detectors are more sensitive to DM than CCD detectors, with the exception of CCD silicon detectors operating with an exposure of $\mathcal{E}=1$~kg-year and a low dark current rate of $\Gamma=1\times10^{-7}$~counts/pixel/day, which exhibit a sensitivity comparable with the one of a HV germanium detector operating with an exposure of $\mathcal{E}=50$~kg-year.~Below $m_\chi=10$~MeV, we find that CCD silicon detectors are more sensitive to DM than germanium and silicon HV detectors, at least for $m_\chi$ as low as 2-3 MeV.~In the best-case scenario, when a HV germanium detector with $\mathcal{E}=50$~kg-year and a CCD silicon detector with $\mathcal{E}=1$~kg-year and $\Gamma=1\times10^{-7}$~counts/pixel/day have simultaneously reported a DM signal, we find that the smallest cross section value compatible with $\mathcal{Z}=3$ ($\mathcal{Z}=5$) is about $8\times10^{-42}$~cm$^2$ ($1\times10^{-41}$~cm$^2$) for $F_{\rm DM}(q^2)=1$, and $4\times10^{-41}$~cm$^2$ ($7\times10^{-41}$~cm$^2$) for $F_{\rm DM}(q^2)=q_{\rm ref}^2/q^2$.

\acknowledgments 
We would like to thank Noah A. Kurinsky and Belina von Krosigk for useful insights into the backgrounds of SuperCDMS. During this work, RC and TE were supported by the Knut and Alice Wallenberg Foundation (PI, Jan Conrad).~RC also acknowledges support from an individual research grant from the Swedish Research Council, dnr. 2018-05029.~The research presented in this article made use of the computer programmes and packages WebPlotDigitizer~\cite{webplotdigitizer}, Wolfram Mathematica~\cite{Mathematica}, Matlab~\cite{MATLAB:2019} and QEdark~\cite{Essig:2015cda}.

\providecommand{\href}[2]{#2}\begingroup\raggedright\endgroup

\end{document}